\documentclass[amsmath,prb,aps,twocolumn]{revtex4}
\usepackage{amsthm,amsfonts,graphicx,verbatim}

\newcommand{\Tr}{{\rm Tr}}

\renewcommand{\cite}[1]{[\onlinecite{#1}]}

\newcommand{\be}{\begin{equation}}
\newcommand{\ee}{\end{equation}}
\newcommand{\bea}{\begin{eqnarray}}
\newcommand{\eea}{\end{eqnarray}}

\newcommand{\la}{\langle}
\newcommand{\ra}{\rangle}
\newcommand{\lb}{\left[}
\newcommand{\rb}{\right]}
\newcommand{\lp}{\left(}
\newcommand{\rp}{\right)}

\renewcommand{\epsilon}{\varepsilon}

\begin{document}
\title{Many-Body Entanglement: a New Application of the Full Counting Statistics}
\author{Israel Klich$^{1}$ and Leonid Levitov$^{2}$}
\affiliation{${}^1$  
Department of Physics, University of Virginia,
Charlottesville VA 22904
\\
${}^2$ Department of Physics, Massachusetts Institute of Technology, Cambridge MA 02139}


\begin{abstract}
Entanglement entropy is a measure of quantum correlations between separate parts of a many-body system, which plays an important role in many areas of physics. Here we review recent work in which a relation between this quantity and the Full Counting Statistics description of electron transport was established for noninteracting fermion systems. Using this relation, which is of a completely general character, we discuss how the entanglement entropy can be directly measured by detecting current fluctuations in a driven quantum system such as quantum point contact.
\end{abstract}

 \maketitle 
\section{Introduction}

Density matrices as a tool for describing quantum-mechanical systems when only partial information about the quantum state is available were introduced in quantum mechanics by Landau in 1927 \cite{Landau}. In this work, called ``The problem of damping in wave mechanics,'' he was interested in irreversibility of certain quantum mechanical processes, such as the spontaneous decay of excited atomic states. Such irreversibility is not inherent to quantum mechanics, it arises from a fully reversible quantum evolution of a larger system, including the variables describing radiation. More generally, Landau was concerned with
the situation when some variables needed to completely describe the system cannot be measured. In such cases the imprecision of our knowledge renders the quantum state vector $\psi$ a useless quantity. Instead, a density matrix must be used, which is defined in the subspace of the system's Hilbert space spanned by those states which can be measured, $\rho=\sum_{ij} \rho_{ij}|i\rangle\langle j|$. 

Another generalization of quantum mechanics in which density matrices feature prominently is quantum statistical mechanics. It was developed, also in 1927, by von Neumann \cite{vonNeumann}, as a way to introduce statistical description in quantum theory. In this approach, a quantum system can occupy states $|i\rangle$, forming an orthonormal set, with statistical probabilities $0\le p_i\le 1$. Such a system is described by $\rho=\sum_{i}p_i|i\rangle\langle i|$, which is nothing but a diagonal representation of the density matrix. In this work, von Neumann first wrote his famous formula for the entropy, 
%
\be\label{eq:vN_entropy}
{\cal S} = -\Tr\lp \rho\log\rho\rp = -\sum_{i}p_i\log p_i
,
\ee
which is a proper extension of the Gibbs entropy (and the Shannon entropy) to the quantum case. For a set of $N$ states, the entropy (\ref{eq:vN_entropy}) can take values varying between $0$ (for a pure state) and $\log N$, realized when all states are occupied with equal probabilities.

\begin{figure}[hb]
\includegraphics*[width=3.0in]{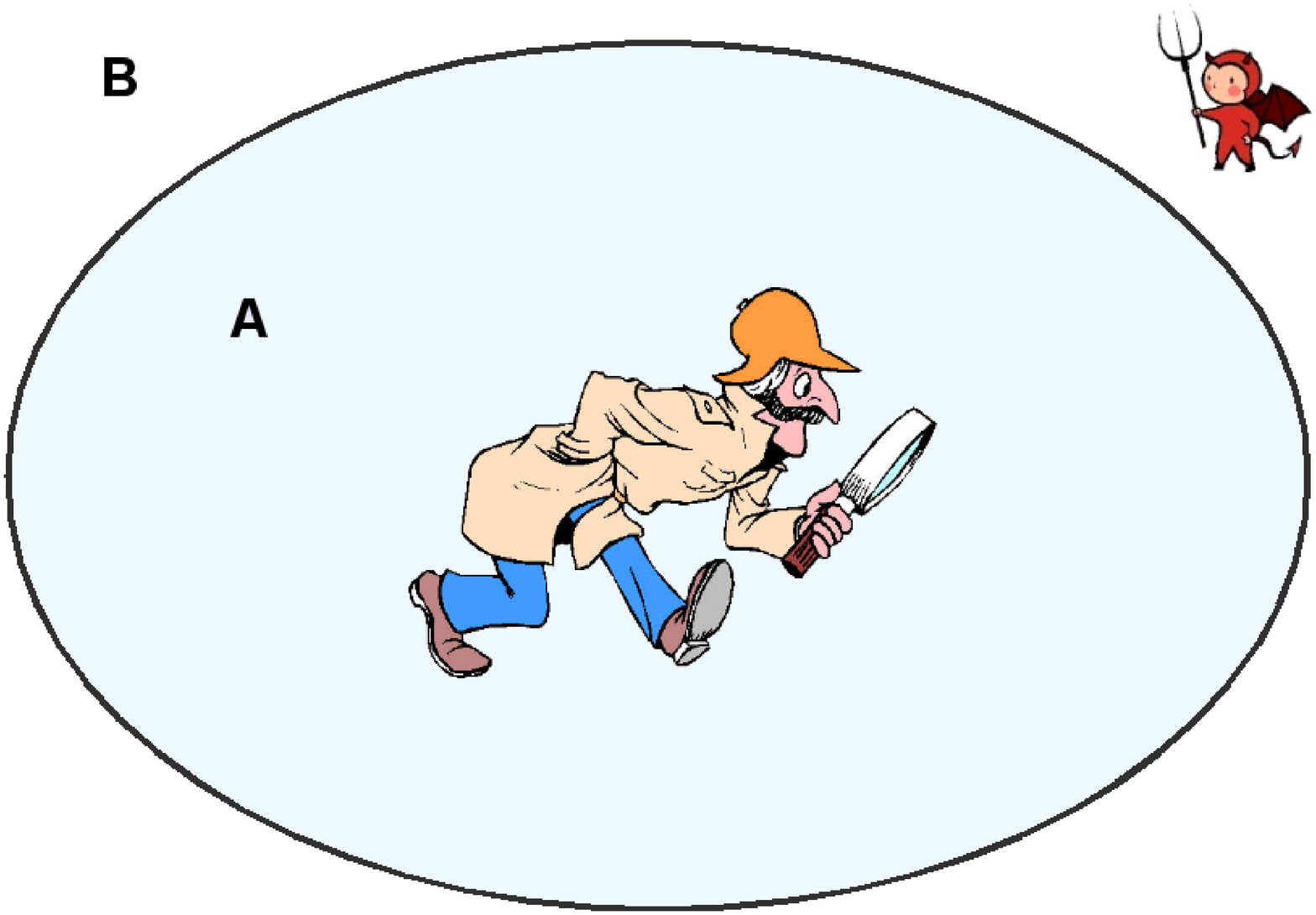}
\caption{
Entanglement between two regions, $A$ and $B$, probed by an observer who can perform measurements only in region $A$ but not in region $B$. The quantum state of the entire system $A+B$, described by the state vector $|0\rangle$, must be ``projected'' on $A$, giving $\rho_A=\Tr_B\rho_0$, where $\rho_0=|0\rangle\langle0|$. Due to entanglement between variables in $A$ and in $B$, the projected state is typically a mixed state, $\rho_A\ne\rho_A^2$, even when the whole system is in a pure state. The von Neumann entropy of the projected state, Eq.(\ref{eq:EE}), is a measure of the entanglement between $A$ and $B$. 
} 
\label{fig1}
\end{figure}

The two aspects of the density matrix, emphasized by Landau and von Neumann, the lack of information about the quantum state in the situation when some variables are not measureable, and the connection with statistical description, are combined in the notion of entanglement entropy. This quantity was introduced in 1986 when Bombelli {\it et al.} \cite{Bombelli86} proposed to use von Neumann entropy, mutual for different parts of a quantum system, as a model of the Bekenstein-Hawking entropy of black holes. They considered a situation when a quantum-mechanical system is described by a pure state which is delocalized between two regions $A$ and $B$ of its configuration space, such that only one of those regions, $A$, is accessible for physical measurement (see Fig.\ref{fig1}). An observer situated in region $A$, after performing a full set of measurments available to him or her, will describe the system by a {\it reduced} density matrix $\rho_A=\Tr_B\rho$, where $\rho=|0\rangle\langle0|$ describes the pure state of the whole system and $\Tr_B$ denotes a trace taken over all variables in region $B$. The von Neumann entropy associated with the projected state, defined as 
\be\label{eq:EE}
{\cal S}_A=-\Tr\, \rho_A\log\rho_A
,
\ee
is a characteristic of entanglement between quantum variables in the regions $A$ and $B$. One can easily check that entanglement entropy does not change when regions $A$ and $B$ are interchanged, $S_B=S_A$. 

These ideas were developed further by Callan and Wilczek \cite{Callan94}, and Holzhey, Larsen and Wilczek \cite{Holzhey94} who considered a $1+1$ dimensional system described by conformal field theory. Taking region $A$ to be an interval of length $\ell$, they found that the entanglment entropy obeys the relation
\be\label{eq:S=1/3logL}
{\cal S}=\frac{c+\bar c}6 \log \frac{\ell}{a}
\ee
where $c$ ($\bar c$) is a central charge of the conformal field theory, and $a$ is a microscopic cutoff length. This finding, which showed that entanglement entropy is sensitive to fundamental characteristics of a quantum system, motivated many further studies in which entanglement entropy  has been used as a tool to analyze many-body states of a variety of different systems \cite{VidalLatorreetal,Refael Moore,Cardy,CalabreseCardyTimeDependentEnt,Bravyi06,EisertOsborne,Cramer08,Bennett96,Verstraete04,Kitaev Preskill,Levin Wen,Fradkin}.

Entanglement entropy, serving as a general characteristic describing quantum many-body correlations between two parts of a quantum system, provided a framework for analyzing quantum critical phenomena \cite{VidalLatorreetal,Refael Moore,Cardy} and quantum quenches~\cite{CalabreseCardyTimeDependentEnt,Bravyi06,EisertOsborne,Cramer08}. Recently it was used as a probe of complexity of topologically ordered states~\cite{Kitaev Preskill,Levin Wen,Fradkin}. In addition, this quantity is of fundamental interest for quantum information theory as a measure of the resources available for quantum computation~\cite{Bennett96} as well
as for numerical approaches to strongly correlated systems~\cite{Verstraete04}.

\section{Measuring the Many-Body Entanglement}

The universal appeal of entanglement entropy arises, at least in part, from the fact that this quantity is defined solely in terms of the many-body density matrix of the system, with no relation to any particular observables whatsoever. 
This is the main reason this quantity has found applications in such diverse fields as cosmology, field theory, condensed matter theory, and quantum information. However, for the very same reason, it has not been clear how to access  this  quantity experimentally, since measuring the entire density matrix of a many-body systems represents a formidable task. Indeed, the many-body density matrix depends on coordinates of all particles in the system, which usually cannot be measured all at once.


Based on what was just said, the very idea of measuring a quantity that encodes information about many-body correlations of a large number of particles, which is what the entanglement entropy is, may seem totally bizarre. Yet, the situation with the entanglement entropy is different from that of the many-body density matrix. Recently we have shown that a direct measurement of entanglement entropy is possibe owing to its relation 
with quantum noise of electric current \cite{KlichLevitov08a}. 

\begin{figure}[hbp]
\includegraphics*[width=3.0in]{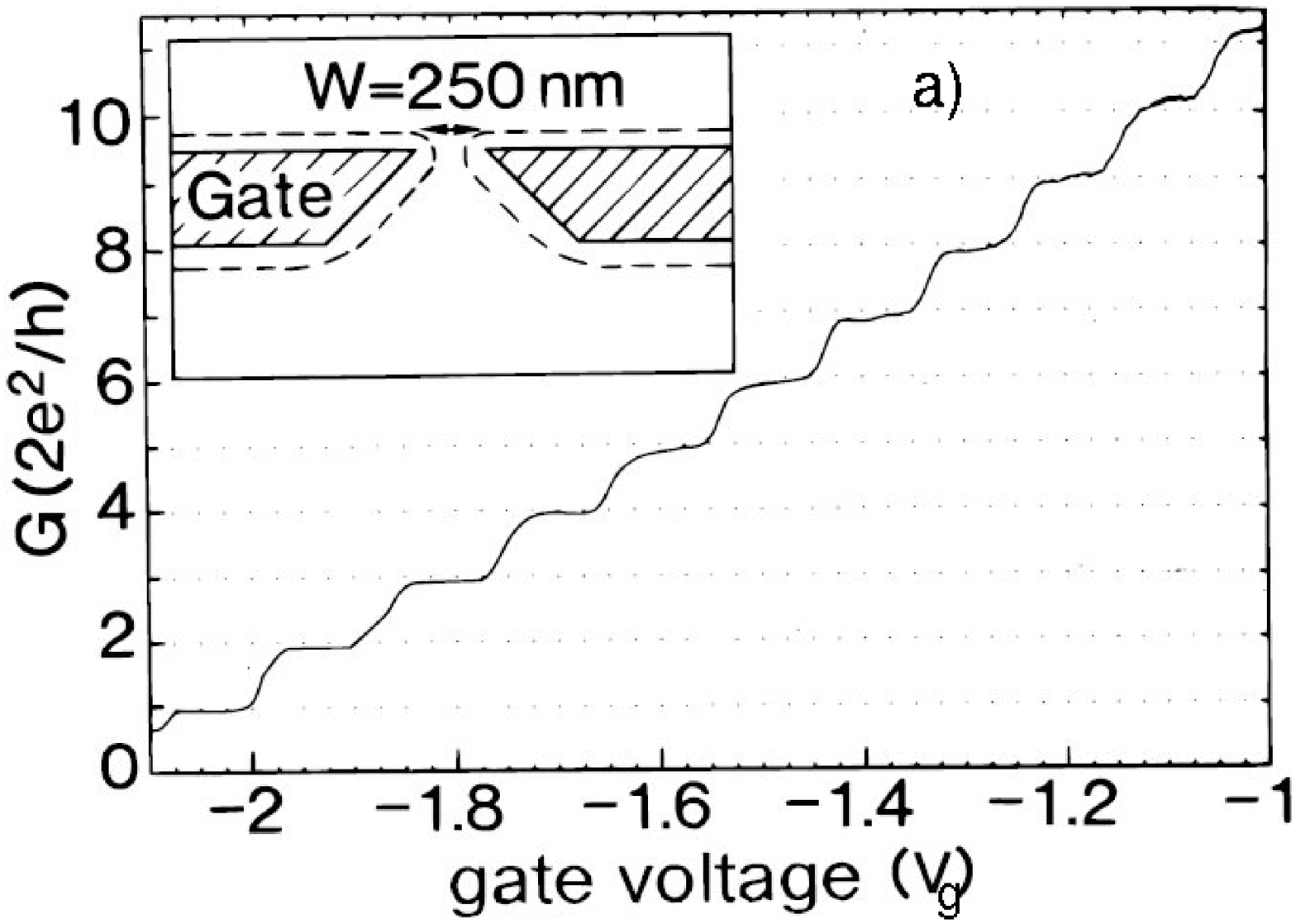}
\includegraphics*[width=3.0in]{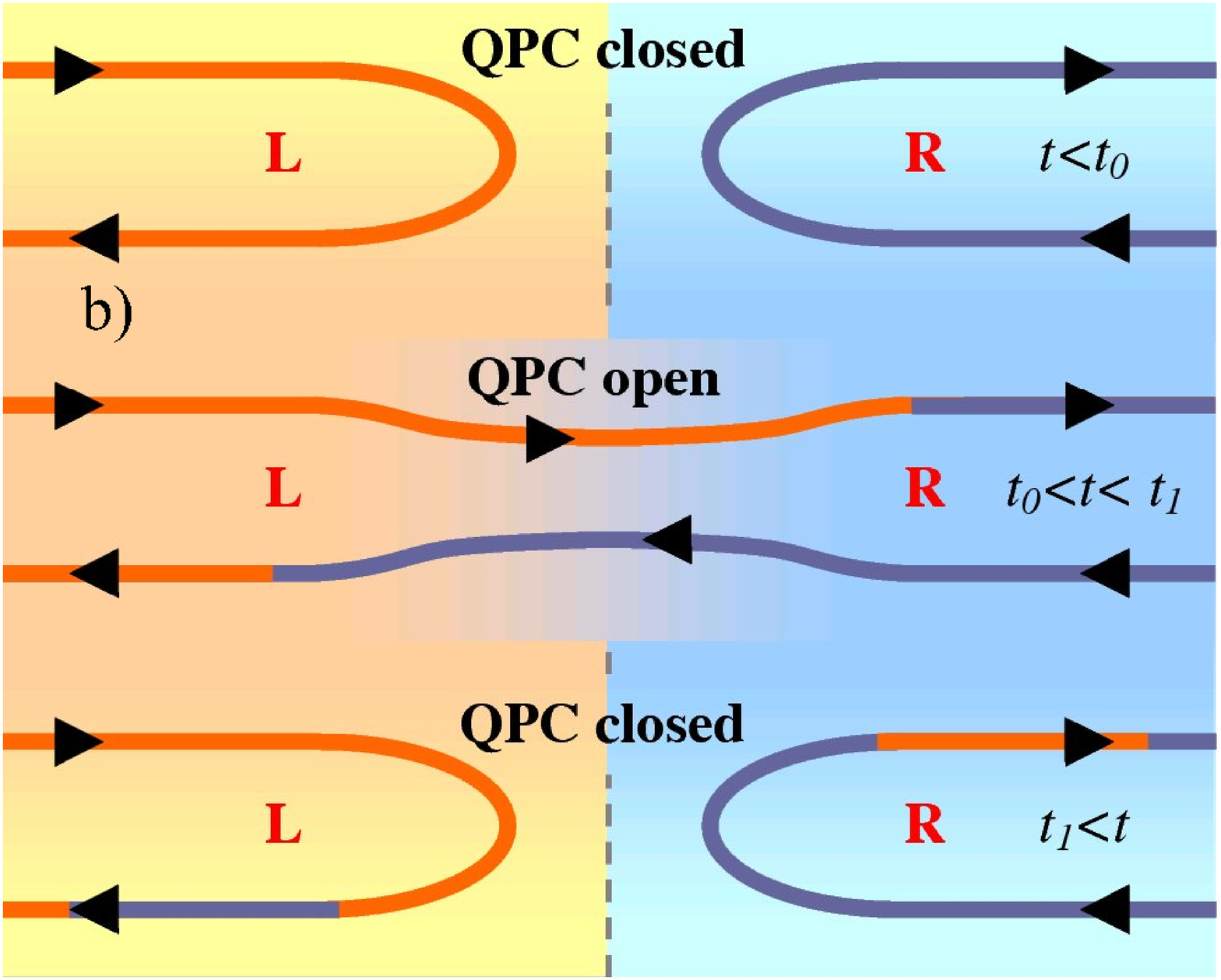}
\caption{
(a) Quantum point contact (QPC), an electron beam-splitter with tunable transmission and reflection (from Ref.\cite{van Wees}). 
Conductance $G$ of a QPC exhibits quantized steps, observed by varying the gate voltage $V_g$. As $V_g$ becomes more negative, the transport channels in the QPC open up one by one, with transmission $D_i$ in channel $i$ increasing from $0$ to $1$ between consecutive steps $i$ and $i+1$. 
(b)
Schematic of two Fermi seas which are connected via a QPC during the time interval $t_0<t<t_1$, and then disconnected (from Ref.\cite{KlichLevitov08a}).
Electron transport, taking place at $t_0<t<t_1$
makes electrons delocalized among the two leads, generating entanglement and current fluctuations. There is a ``space-time'' duality between this situation and the conventional approach~\cite{Bombelli86,Callan94,Holzhey94},
in which many-body correlations are analyzed using a finite region in space.} 
\label{fig2}
\end{figure}

The system analyzed in Ref.\cite{KlichLevitov08a} 
is a quantum point contact, representing an electron beam-splitter with transmision and reflection coefficients tunable by external gates \cite{van Wees}. In essence, the QPC serves as a door between electron reservoirs, which can be opened and closed on demand. By adjusting the voltage on the gates of the QPC, different transmission channels in it can be opened or closed individually, leading to quantized steps in the QPC conductance (see Fig.\ref{fig2}a).  

The simplest protocol of driving the QPC that leads to entanglement of many-body states in the reservoirs is illustrated in Fig.\ref{fig2}b. We start with the QPC in a closed state, when the resrvoirs are disconnected, then the QPC is opened during time $t_0<t<t_1$, and then closed again. In this process, most of the particles either remain in their initial resrvoir, or are fully transmitted to another reservoir. Some particles, however, become delocalized between the two reservoirs, making the states in the two reservoirs entangled. 
After the QPC is closed any communication between the reservoirs becomes impossible. This situation mimics that considered in the definition of many-body entaglement, when an observer can perform measurements only in one region but not in the other (see Fig.\ref{fig1}). 

The state of two Fermi seas, coupled via the QPC as shown in Fig.\ref{fig2}b, evolves as a pure state until projection on a specific reservoir is performed. After projecting on reservoir $L$ at $t=t_1$, the density matrix takes the form
%
\be\label{eq:evolution+projection}
  \rho_L(t_1)=\Tr_R ({\bf U}(t_1,t_0)\rho_0{\bf U}^{\dag}(t_0,t_1))
,
\ee
where $\rho_0$ is the initial state,
${\bf U}$ is the many-body evolution between $t_0$ and $t_1$, and $Tr_R$ is a partial trace over degrees of freedom in the lead $R$. Due to the exchange of particles between reservoirs during $t_0<t<t_1$, the resulting density matrix describes a mixed state with a nonzero von Neumann entropy
\be\label{eq:S_L}
{\cal S}_L=-\Tr\rho_L(t_1)\log\rho_L(t_1)
\ee
which characterizes entanglement buildup due to particle exchange between reservoirs.

The evolution of two coupled Fermi seas, describing this process, was analyzed in Ref.\cite{KlichLevitov08a}. This can be done exactly owing to the free fermion nature of the problem. It was found that all multi-particle correlations in the Fermi sea that are relevant for entanglement are fully accounted for by temporal correlations of electric current flowing through the QPC. Specifically, there is a universal relation between the entanglement entropy \eqref{eq:S_L} and the full counting statistics of the charge transmitted through the QPC. This relation, which we shall discuss below, is of a completely general nature, independent of the details of the protocol used to drive the QPC. As such it can be used to obtain the entanglement entropy from measured fluctuations of electric current.

The relation between entanglement and electric noise has been at the center of the discussion of different ways to generate entangled pairs in a driven electron system, using transport in normal metal-superconductor junctions \cite{Lesovik01,Chtchelkatchev02} and in the QPC \cite{Beenakker03,Samuelsson03}. Such pairs, which represent an electron analog of the recently demonstarted entangled photon pairs \cite{Walborn}, could be used for testing Bell inequalities in a condensed matter system.

In contrast to Refs.\cite{Lesovik01,Chtchelkatchev02,Beenakker03,Samuelsson03}, here we are concerned with entanglement of many-body states, represented by Fermi seas in the right and left reservoirs shown in Fig.\ref{fig2}b. This entanglement is generated by the evolution of the full many-body fermion state happening when the QPC is opened and then closed. In that, there is an analogy with recent literature in which generation of entanglement in time for critical Hamiltonians~\cite{CalabreseCardyTimeDependentEnt} and for generic Hamiltonians~\cite{Bravyi06,EisertOsborne} was discussed. 

The centerpiece of the approach of Ref.\cite{KlichLevitov08a} is the relation between many-body entanglement and a physical measurement, which in this case is electric current fluctuations. A relation of entanglement with another measurable quantity, particle number statistics, was emphasized in Ref.\cite{KlichRefaelSilva}. In this paper a Fermi system with a fixed total number of particles was considered in a setup pictured in Fig.\ref{fig1}. The many-body entanglement in this system was expressed through the probability distribution of particle number in the region $A$. Similarly, in the approach discussed here, we link the entanglement generated in a driven fermion system, such the QPC, to the statistics of charge transmitted between reservoirs. 

The relation between entanglement and counting statistics of charge fluctuations, which we discuss below, was derived in Ref.\cite{KlichLevitov08a} for a noninteracting fermion system. In the derivation we focus on the QPC as a convenient model, however it will be clear that the result is more general. The overall simplicity of the relation between entanglement and counting statistics, and also its independence of the details of the driving protocol, suggests an even higher degree of generality. It would be extremely interesting to find out whether a similar relation holds for {\it interacting} many-body systems, such as quantum spin chains, Luttinger liquids or Quantum Hall liquids. 








\section{A Primer on Counting Statistics}

The Full Counting Statistics (FCS) approach has been developed in the theory of quantum noise to describe current fluctuations in nanodevices such as QPC and tunnel junctions~\cite{LevitovLesovik}. These fluctuations can be characterized by the probability distribution of charge transmitted through the device during the measurment. It is convenient to combine individual probabilities in a single quantity, the generating function 
\be\label{eq:chi_Pn}
\chi(\lambda)=\sum_{n=-\infty}^\infty P_ne^{i\lambda n},
\ee 
where $P_n$ is the probability to transmit $n$ charges. The auxiliary variable $\lambda$ is sometimes called ``counting field'' in the literature. 

For example, a binomial distribution with the number of attempts $N$ and the probabilities to succeed and fail in each attempt $p$ and $q=1-p$ is described by
%
\be\label{eq:binomial}
P_n=\lp\begin{array}{c} N \\ n\end{array}\rp p^nq^{N-n}
,\quad
\chi(\lambda)=\lp 1-p+pe^{i\lambda}\rp^N
.
\ee
Probability distribution of this form arises in the problem of a DC-biased QPC \cite{LevitovLesovik}.

The function $\chi(\lambda)$ encodes all cumulants of FCS (or, irreducible moments) via an expansion
\be\label{eq:chi_Cn}
  \log\chi(\lambda)=\sum_{m=1}^\infty {(i\lambda)^mC_m\over m!}
.
\ee
The lowest cumulants $C_1$, $C_2$, $C_3$... describe properties of the distribution $P_n$ such as the mean $\bar n$, the variance $\la (n-\bar n)^2\ra$, the skewness $\la (n-\bar n)^3\ra$, etc. 

\begin{figure}[hbp]
\includegraphics*[width=3.0in]{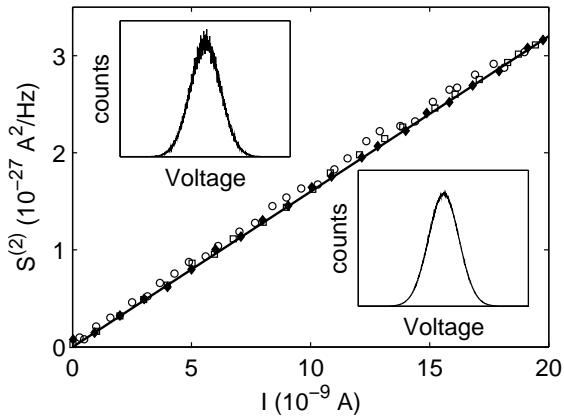}
\caption{
Measurement of high-order cumulants of electron shot noise in a tunnel junction
(from Ref.\cite{BomzeReznikov}). Fluctuations of electric current, integrated over a short time interval $\Delta t=5\,{\rm ns}$ and histogrammed, give the probability distribution of transmitted charge (insets). The mean, the variance and the skewness of this distribution are used to obtain the cumulants of FCS, $C_1$, $C_2$ and $C_3$. The noise power $S_2$, obtained from $C_2$, agrees with the expectation for Poisson statistics, $C_2=C_1$ (black line). The measured value of the 3rd cumulant is also close to the poissonian value, $C_3\approx C_1$ (not shown, see Fig.3 in Ref.\cite{BomzeReznikov}). Higher-order cumulants are challenging to obtain because the histogram of transmitted charge is nearly gaussian for nanosecond sampling times and currents of a few nA, as expected from the central limit theorem.}
\label{fig3}
\end{figure}

The 2nd cumulant $C_2$ is available from routine noise measurement. 
Recently, the 3rd cumulant $C_3$ has been measured in tunnel junctions~\cite{Reulet03,BomzeReznikov} and in QPC~\cite{Gershon07}, while
cumulants up to 5th order where measured in quantum dots~\cite{Fujisawa,Gustavsson}. In fact, the method used in Refs.\cite{BomzeReznikov,Gershon07} yields the full probability distribution $P_n$ (see Fig.\ref{fig3}); however, only the lowest moments $C_1$, $C_2$ and $C_3$ of this distribution were found to be dominated by intrinsic effects.

Theoretical description of electron transport in a QPC involves scattering states constructed from the transmission and reflection amplitudes $A(t)$, $B(t)$, which in general are time-dependent. In a Schr\"odinger representation, the evolution of wavepackets undergoing scattering between the left and right reservoirs is described by
%
\be\nonumber
U\begin{bmatrix}
 |x\ra_L  \\
 |x\ra_R \\
\end{bmatrix}
=
\begin{bmatrix}
  B(t_x) & A(t_x) \\
  -A^*(t_x) & B^*(t_x) \\
\end{bmatrix}
\begin{bmatrix}
 |x(t)\ra_L  \\
 |x(t)\ra_R \\
\end{bmatrix}
,\quad
x<0<x(t)
,
\ee
and $U |x\ra_{L,R}=|x\ra_{L,R}$ otherwise.
Here $x(t)=x+v_Ft$ is the guiding center coordinate of a wavepacket,
$t_x=-x/v_F$ is the time of arrival at the scatterer, $v_F$ is the Fermi velocity,
and $|x\ra_{L,R}$ describes incoming ($x<0$) and outgoing ($x>0$) wavepacket states in the leads.

Crucially, the FCS generating function (\ref{eq:chi_Pn}) can be expressed through the evolved many-body density matrix of the initial state, projected on one of the leads after the evolution is completed, as in Eq.\eqref{eq:evolution+projection}. This relation, derived in Refs.\cite{Abanov Ivanov,KlichLevitov08a}, is outlined below, and then used to find the entanglement entropy.


There are certain general properties of the evolved density matrix in a noninteracting fermion system which are best understood by considering evolution of a gaussian state 
\be\label{eq:gaussian}
\rho=\frac1{Z}\exp\lp -\sum_{ij}H_{ij}a_i^\dagger a_j\rp
,\quad
Z=\det(e^{-H}+1)
,
\ee
where $H$ is a general hermitian operator in the single-particle Hilbert space, and $Z$ is the normalization factor. The state of interest, describing the QPC at zero temperature, represents a particular case of Eq.(\ref{eq:gaussian}). 

One property which greatly simplifies the analysis, is that the state (\ref{eq:gaussian}) remains gaussian under the evolution (\ref{eq:evolution+projection}). This follows from the observation that the Schr\"odinger evolution of $a_i$'s is equivalent to the single-particle Heisenberg evolution of $H_{ij}$. Gaussian form of the state is also preserved under projection \cite{KlichLevitov08a}.

This property can be used to reduce the many-body quantities of interest to certain one-particle quantities. Indeed, any gaussian state (\ref{eq:gaussian}) can be described by a matrix in the single-particle Hilbert space defined as
\be
n_{ij}=\Tr\lp \rho a_i^\dagger a_j\rp=\lb \lp e^H+1\rp^{-1}\rb_{ij}
.
\ee
In particular, for a fully filled Fermi sea in both reservoirs the matrix $n$ is a projector on the subspace of all states with negative energy, $\epsilon_L<0$, $\epsilon_R<0$. As a projector, the matrix $n$ satisfies the relation $n^2=n$. 

In what follows, we will need to consider evolution of the projector $n$,
followed by projection on the left reservoir $L$, 
which is described by
\be\label{eq:M_definition}
n_U=UnU^\dagger
,\quad
M=P_Ln_UP_L
,
\ee
where $P_L$ is a projection on the modes in $L$. The matrix $n_U$ is a projector  describing evolved Fermi sea, whereas the matrix $M$, which is of main interest for us, is given by a product of three projectors. Thus generally $M$ is not a projector.


To illustrate the time evolution of single-particle quantities, such as $n_U$, we recall that in the FCS approach it is convenient to work in a time representation~\cite{LevitovLesovik04}, labeling states by times of arrival at the scatterer $t_x$. In this representation 
the initial Fermi projection is given
by $n(t,t')={1\over 2\pi i(t-t'+i0)}I$ with $I$
 a $2\times 2$ identity matrix in the $L$, $R$ basis. The evolved state $n_U$ is given by
\begin{eqnarray}\nonumber
n_U(t,t')=U(t)n(t,t')U^{\dag}(t')
,\quad
U(t)=
\begin{bmatrix}
  B(t) & A(t) \\
  -A^*(t) & B^*(t) \\
\end{bmatrix}
.
\end{eqnarray}
%
The evolution operator is diagonal with respect to the arrival time label $t$, which is precisely why this representation is so convenient. 

The FCS generating function \eqref{eq:chi_Pn} can be expressed in terms of the single-particle quantities, such as $n$, $n_U$, $P_L$ and $M$, in several different ways. The first representation of this kind, found in Ref.\cite{LevitovLesovik}, involves a functional determinant
%
\be\label{FCS}
   \chi(\lambda)=
 \det(1-n +n  U^{\dag}e^{i\lambda P_L}U e^{-i\lambda
  P_L})
.
\ee
%
This determinant, which must be properly regularized for infinitely deep Fermi sea \cite{MuzukantskiiAdamov,Avron Graf}, can be explicitly evaluated, yielding the FCS for various driving protocols of the QPC~\cite{Ivanov93,Levitov96,Ivanov97}.

Another useful representation of the functional determinant giving $\chi(\lambda)$ was obtained recently in Ref.\cite{Abanov Ivanov} (a similar relation was derived in a related problem of particle number fluctuations \cite{Klich06,KlichRefaelSilva}), 
\be\label{eq:chi-M}
\chi (\lambda)=\det\Big((1-M +M 
e^{i\lambda })e^{-i\lambda (n P_L)_U}\Big)
,
\ee
where $(n P_L)_U=Un P_LU^\dagger$ (see discussion in \cite{KlichLevitov08a}). Here $M$ is the Fermi sea in $L$, evolved by $U$ and projected back to $L$ by $P_L$ (see Eq.\eqref{eq:M_definition}). The unitary operator $e^{-i\lambda (n P_L)_U}$ contributes a multiplicative factor of the form $e^{ix\lambda}$ to the determinant, which may only affect the first cumulant $C_1$ of the FCS generating function, Eq.(\ref{eq:chi_Cn}). 

The representation (\ref{eq:chi-M}) is of interest because it reveals certain general features of $\chi(\lambda)$. It is convenient to introduce the spectral density of $M$, defined by $\mu(z)=\Tr\, \delta(z-M)$. Since $M$ is a product of three projectors, $M=P_Ln_UP_L$, all its eigenvalues lie in the interval $0\le z\le 1$ (see Refs.\cite{Vanevic,Abanov Ivanov,Sherkunov}). Using the spectral density, we can rewrite (\ref{eq:chi-M}) as
\be
\log\chi(\lambda)=ix\lambda+\int_0^1 dz\mu(z)\log\lp 1-z+ze^{i\lambda }\rp
.
\ee
%
This expression indicates that the FCS always assumes a generalized binomial form, $\chi(\lambda)\propto \prod_z \lp 1-z+ze^{i\lambda }\rp^{\mu(z)}$, with the product taken over the entire spectrum of $M$.

\section{The relation between entanglement and Full Counting Statistics}

The entanglement generated as a result of evolution of two Fermi seas coupled through a QPC is characterized by the von Neumann entropy of the density matrix 
\eqref{eq:evolution+projection}, ${\cal S}=-\Tr\, \rho_L(t_1)\log\rho_L(t_1)$. As we discussed above, this matrix is of a gaussian form, Eq.\eqref{eq:gaussian}. This property can be used to express entropy through single-particle quantities (see Refs.\cite{Peschel,KlichLevitov08a}), giving
\be\label{EntropyM}
{\cal S}=-\Tr\, \lp M\log M+(1-M)\log(1-M)\rp
\ee
where $M$ is defined in \eqref{eq:M_definition}, and the trace is taken in
the space of single-particle modes in $L$.  

If the spectral density of $M$ is known, the entropy \eqref{EntropyM} can be written as
\be\label{eq:S_mu(z)}
{\cal S}=-\int_0^1dz\mu(z)\lp z\log z+(1-z)\log(1-z)\rp
\ee
At the same time, as we discussed above, the FCS generating function is also expressed through the spectral density of $M$. Furthermore, the spectral density is encoded in the generating function. This can be seen most easily by rewriting Eq.\eqref{eq:chi-M} as
\be\label{eq:chi(z)}
\chi (z)=\det\lp (z-M) e^{-i(n P_L)_U\lambda(z)}(1-e^{i\lambda(z)})\rp
,
\ee
where the parameter $\lambda$ was changed to $z=(1-e^{i\lambda})^{-1}$. Thus the resolvent of $M$ is given by the derivative $\partial_z\log \chi(z-i0)$ up to a sum of two terms $\frac{a_0}{z}+\frac{a_1}{z-1}$ arising from the last two factors in \eqref{eq:chi(z)}.

Using this observation the entanglement entropy can be expressed through the FCS cumulants $C_m$ 
\cite{KlichLevitov08a}. Writing $\mu(z)$ in Eq.\eqref{eq:S_mu(z)} as $\frac1{\pi}{\rm Im}\,\partial_z\log\chi(z-i0)$ and plugging into it the relation \eqref{eq:chi_Cn}, we integrate over $z$ to obtain
%
\begin{eqnarray}\label{Entropy from cumulants}
{\cal S}=\sum_{m>0} {\alpha_m\over  m !}C_{m} 
,\quad
\alpha_m=
\Big\{
\begin{array}{cc}
    (2\pi)^m |B_{m}|, &  $m$\,\,{\rm even} \\
  0,  &  $m$\,\,{\rm odd} 
\end{array}
,
\end{eqnarray}
where $B_m$ are Bernoulli numbers, $B_2=\frac16$, $B_4=-\frac1{30}$, $B_6=\frac1{42}$... These numbers are defined by the generating function $\frac{x}{e^x-1}=\sum_{n=0}^{\infty}B_n {x^n\over n!}$. Interestingly, only even cumulants contribute to the entropy. The first few terms in the series \eqref{Entropy from cumulants} are:
\begin{eqnarray}\label{Entropy C2 C4 C6}
  {\cal S}={\pi^2\over 3} C_2+{\pi^4\over 15}C_4+{2\pi^6\over 945}C_6+...
\end{eqnarray}
Asymptotically, $|B_{n}|\approx{2\, n!\over (2\pi)^{n}}$ for large even $n$,
which means that the coefficients in (\ref{Entropy from cumulants}) 
stay bounded, $\alpha_m/m!\approx 2$ for large $m$.

It is clear from this derivation that the relations \eqref{Entropy from cumulants} and \eqref{eq:S_mu(z)} are completely general and valid for arbitrary driving. The formula \eqref{Entropy from cumulants}  can be used to determine the entanglement entropy from the values of FCS moments, whereas the formula \eqref{eq:S_mu(z)} can be used when the spectral density $\mu(z)$ is known. 

As an example illustrating these relations, it is instructive to consider a QPC biased with a DC voltage. The FCS for a DC-biased QPC with constant transmission $0\le D\le 1$ is known to have binomial form \cite{LevitovLesovik}, described by the generating function \eqref{eq:binomial},
\be
\chi(\lambda)=\lp 1-D+De^{i\lambda}\rp^N
,\quad
N=eV\Delta t/h
,
\ee
where $N$, given by the product of the bias voltage $V$ and the measurement time $\Delta t=t_1-t_0$, is interpreted as the ``number of attempts.'' Comparing this expression with \eqref{eq:chi-M}, we infer that the spectral density of $M$ in this case is a delta function, $\mu(z)=N\delta(z-D)$. Plugging this result in the expression  \eqref{eq:S_mu(z)}, we find that entanglement entropy is generated at a constant rate given by
\be\label{eq:dS/dt_DC}
d{\cal S}/dt=-\lp D\log D+(1-D)\log(1-D)\rp eV/h
\ee
Production of entanglement in a DC-biased QPC was considered in Ref.\cite{Beenakker_dS/dt}, were the result \eqref{eq:dS/dt_DC} was obtained. The rate of entanglement production is zero for $D=0,1$ and maximal for $D=1/2$.

In this case, using the series \eqref{Entropy from cumulants} turns out to be not too convenient because of a large number of high-order cumulants that contribute to the result. This can be seen most directly in the limit of small $D\ll1$, when the entanglement production rate scales as $D\log(1/D)$, while the cumulants $C_m\sim D$. This means that there are about $\log(1/D)$ terms in the series \eqref{Entropy from cumulants} giving  contributions of the same order of magnitude.

\section{Connecting and disconnecting Fermi seas}



Here we shall discuss the protocol of driving the QPC in which it is opened at $t_0<t<t_1$ and closed at $t<t_0$ and $t_1<t$, as illustrated in Fig.\ref{fig1}b. In the simplest case considered here, the QPC is unbiased, i.e. the right and left reservoirs remain at equal chemical potentials at all times. The fluctuations of charge transmitted through the QPC in this case are gaussian \cite{LevitovLesovik04}, 
\be\label{eq:chi_eta(lambda)}
\chi(\lambda)=e^{-\frac12\lambda^2C_2}
,\quad
C_2=\frac1{\pi^2}\log\frac{\Delta t}{\tau}
,
\ee
where $\Delta t=t_1-t_0$ is the time window during which the QPC was open, and $\tau$ is a short-time cutoff of order of the on/off switching time.

The simplest way to estimate entanglement production is to use the formula \eqref{Entropy from cumulants}. Since for a gaussian distribution all cumulants are zero except $C_2$, we find
\be\label{eq:S=1/3logT}
{\cal S}={\pi^2\over 3} C_2=\frac13\log\frac{\Delta t}{\tau}
.
\ee
This result resembles the logarithmic dependence predicted by conformal field theory,
Eq.\eqref{eq:S=1/3logL}. In fact, since for free fermions the central charge is $c=\bar c=1$, the prefactor before the logarithm in Eq.\eqref{eq:S=1/3logL} is the same as in Eq.\eqref{eq:S=1/3logT}.

The similarity between the results \eqref{eq:S=1/3logT} and \eqref{eq:S=1/3logL} for entanglement entropy is of course not accidental. From a field-theoretic viewpoint, space and time play the same role in a conformal field theory. Therefore, analyzing entanglement using a window of size $\ell$ in space should be equivalent to doing it using a window of size $\Delta t=\ell/v_F$ in time, where $v_F$ is Fermi velocity. This is precisely what the comparison of ``time-like'' Eq.\eqref{eq:S=1/3logT} and ``space-like'' Eq.\eqref{eq:S=1/3logL} suggests.

One can also understand the relation between the results \eqref{eq:S=1/3logT} and \eqref{eq:S=1/3logL} in a more intuitive way, without relying on a space-time duality. For that, we consider in more detail the process of mixing of two Fermi seas shown in Fig.\ref{fig2}b. Using different colors (blue and red) to mark particles in different reservoirs, we observe that after $t=t_2$, when the reservoirs are disconnected, there is a group of blue particles in the red Fermi sea. Simultaneously, there is a group of red particles in the blue Fermi sea. 

\begin{figure}[hbp]
\includegraphics*[width=3.0in]{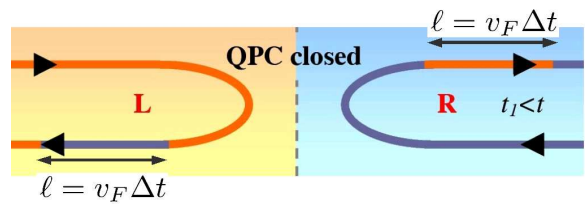}
\caption{Entanglement of two Fermi seas generated by connecting and disconnecting them via a QPC during time interval $\Delta t$ can be interpreted as entanglement in a stationary Fermi sea probed with a window of size $\ell=v_F\Delta t$.
}
\label{fig4}
\end{figure}

Assuming, without loss of generality, ballistic dynamics in each of the leads with constant velocity $v_F$, we find that the blue and red groups of particles occupy spatial regions of size $\ell=v_F\Delta t$.
Since there is no correlation between the left and right reservoirs in the initial state of the system, we conclude that in the final state, shown in Fig.\ref{fig4}, there is no correlation between blue and red particles either, even if they reside in the same reservoir. This means that the von Neumann entropy evaluated for one of the reservoirs, in which both red and blue particles are present simultaneously, will be the same as the entanglement entropy found for a single Fermi sea with a window of size $\ell=v_F\Delta t$.


This argument can in fact be made rigorous using the method of Ref.\cite{LevitovLesovik04}. In this paper, concerned with the FCS of the process shown in Fig.\ref{fig2}b, the problem of fluctuations of charge transmitted through the QPC during a time interval $\Delta t$ was mapped on the problem of particle number fluctuations in an interval of size $\ell=v_F\Delta t$. Using the bosonization representation of a free Fermi gas, the latter fluctuations can be shown to be gaussian, giving Eq.\eqref{eq:chi_eta(lambda)}.

The result \eqref{eq:S=1/3logT} can be readily generalized to more complicated protocols of driving the QPC. In particular, it interesting to consider the QPC switching between the on and off states multiple times $t^{(1)}_0<t^{(1)}_1<...<t^{(N)}_0<t^{(N)}_1$. In this case, generalizing the above argument, we find gaussian charge statistics
\bea\label{eq:chi_multiple}
&& \chi(\lambda)=e^{-\frac12\lambda^2 C_2}
,\quad
C_2=\frac1{2\pi^2}G
,
\\\label{eq:G_N}
&& G =\sum_{i, j=1}^N\log\frac{t_1^{(i)}-t_0^{(j)}}{t_0^{(i)}-t_0^{(j)}}
+\log\frac{t_1^{(i)}-t_0^{(j)}}{t_1^{(i)}-t_1^{(j)}}
,
\eea
Using the relation \eqref{Entropy from cumulants} with the only nonvanishing contribution due to $C_2$, we obtain the entropy
\be\label{eq:S=1/3G}
{\cal S}={\pi^2\over 3} C_2=\frac16 G
.
\ee
The case of multiple switching provides a time-like realization of the situation studied in Ref.\cite{Cardy}, where entanglement of a conformal field theory was analyzed using a system of non-overlapping windows $x^{(1)}_0<x^{(1)}_1<...<x^{(N)}_0<x^{(N)}_1$. As in the case of single switching, Eq.\eqref{eq:S=1/3logL}, the answers for the entanglement entropy coincide after distances are converted to times via $x_{0,1}^{(j)}=v_Ft_{0,1}^{(j)}$.

Multiple switching of the QPC between the on and off states, repeated periodically in time, can be used to convert the current fluctuations, described by Eqs.\eqref{eq:chi_eta(lambda)},\eqref{eq:chi_multiple}, into the DC shot noise which can be measured by conventional techniques. In Ref.\cite{KlichLevitov08a}, the effective temperature of such noise was estimated to be about $25\,{\rm mK}$ for the driving frequency $\nu=500\,{\rm MHz}$, putting it in the experimentally feasible range.  

In view of the possibility of such an experiment, it is interesting to consider how the above results are changed if the QPC transmission in the open state is less than one. This problem can be readily addressed both for single and multiple switching protocols, since the FCS in this case has been known \cite{LevitovLesovik04},
\be\label{eq:chi_eta(lambda_*)}
\chi(\lambda)=e^{-\frac{\lambda^2_\ast}{4\pi^2}G}
,\quad
\sin\frac12\lambda_\ast=\sqrt{D}\sin\frac12\lambda
,
\ee
with $G$ given by (\ref{eq:G_N}) as above, and 
$D<1$ the QPC transmission coefficient. 

Notably, in this case $\chi(\lambda)$ is non-gaussian. Thus the simplest way to find the entropy is to use its relation with the spectral density of $M$, Eq.(\ref{eq:S_mu(z)}). The spectral density $\mu(z)$ can be evaluated using the resolvent $\Tr \lp 1/(z-M-i0)\rp$, which is found from $\chi(\lambda)$, Eq.\eqref{eq:chi_eta(lambda_*)}, as discussed above. This gives a peculiar function \cite{KlichLevitov08a} which vanishes inside the interval
\[
z_1<z<z_2
,\quad
z_{1,2}=\frac12\lp 1\mp\sqrt{1-D}\rp
,
\]
while outside this interval, at $0<z<z_1$ and $z_2<z<1$, it is given by
\be\label{eq:mu(z)_D}
\mu(z)=\frac{G}{2\pi^2}\frac{|1-2z|\sqrt{D}}{z(1-z)\sqrt{(z-z_1)(z-z_2)}}
.
\ee
The entropy, found from 
\eqref{eq:S_mu(z)}, exhibits a logarithmic dependence on the times $t_{0,1}^{(i)}$ identical to (\ref{eq:G_N}); the only change is a $D$-dependent prefactor. 

The behavior of the factor $F={\cal S}(D)/{\cal S}(1)$ describing entropy reduction due to imperfect transmission in the QPC was analyzed in Ref.\cite{KlichLevitov08a}. It was found that, unless $D$ is very small, the reduction in entropy can be attributed mostly to the change in the second cumulant, $C_2=\frac{D}{2\pi^2}G$, with the contribution from higher cumulants being relatively small. Thus even for imperfect QPC transmission, the DC shot noise generated by QPC switching gives a reasonably good estimate of the entanglement entropy production.

\section{Conclusions}

There are several ways in which the relation between  many-body entanglement and the Full Counting Statistics (FCS) description of quantum transport comes as noteworthy and surprising. First, it provides an interesting new application for the FCS approach. Using the framework outlined above, many results from the FCS literature can be reinterpreted and used to study entanglement production in a variety of regimes of experimental interest. 

Second, the FCS approach offers new insight into the nature of many-body entanglement in driven quantum systems. In particular, two different kinds of entanglement, called restricted and unrestricted, are distinguished in quantum information theory \cite{Beenakker_dS/dt}. This distinction refers to our ability to detect entanglement by means of local measurements, as required in various teleportation and cryptography protocols. Entanglement production in a system such as the QPC, opened and closed multiple times as discussed above, results from particle exchange between two Fermi seas. However, since particle number is a conserved quantity, some of the produced entanglement may be inaccessible to local measurements, and thus unuseful from a quantum information standpoint. 

How large is the unrestricted (``useful'') entanglement? This question can be addressed by generalizing the FCS-based approach, as discussed in detail in Ref.\cite{KlichLevitov08b}. It turns out that for realistic driving protocols, such as those analyzed above, nearly all entanglement is unrestricted. Specifically, when the number of particles transferred through the QPC becomes large, the restricted entanglement entropy scales as $\log{\cal S}$, where ${\cal S}\gg 1$ is the total entanglement entropy. Thus only a small fraction of entanglement is degraded to the restricted form due to particle conservation.

Finally, the relation between entanglement and FCS opens a way to perform direct measurement of entanglement entropy by detecting fluctuations of electric current in a driven system. In particular, by using the QPC switching periodically between the on and off states, and utilizing space-time duality of one-dimensional systems, the relation between entanglement production and noise can be used to test the seminal ${\cal S}=\frac13 \log L$ prediction of conformal field theory~\cite{Holzhey94}. More generally, this relation offers a method for theoretical  and experimental investigation into the nature of many-body entanglement, and in particular, of its build up in non-equilibrium quantum systems.

We thank  C. W. J.  Beenakker, B. Reulet, and M. Reznikov for useful discussions. This research was supported in part by US-Israel Binational Science Foundation and W. M. Keck Foundation
Center for Extreme Quantum Information Theory (L.L.).




\begin{thebibliography}{99}


\bibitem{Landau}
L. D. Landau, Z. Phys. {\bf 45}, 430 (1927).

\bibitem{vonNeumann}
J. von Neumann, 
G\"ott. Nach. {\bf 1}, 273 (1927).

\bibitem{Bombelli86}
L. Bombelli, R. K. Koul, J. Lee, and R. D. Sorkin, Phys. Rev. D{\bf 34} 373 (1986).

\bibitem{Callan94} C. Callan, F. Wilczek, Phys. Lett. B{\bf 333}, 55 (1994)

\bibitem{Holzhey94}
C. Holzhey, F. Larsen and F. Wilczek, Nucl. Phys. B{\bf 424}, 443 (1994).


\bibitem{VidalLatorreetal}
G. Vidal, J. I. Latorre, E. Rico and A. Kitaev, Phys. Rev. Lett.
{\bf 90}, 227902 (2003).


\bibitem{Refael Moore}
G. Refael and J. E. Moore, Phys. Rev. Lett. {\bf 93} 260602
(2004).

\bibitem{Cardy}
P. Calabrese and J. Cardy, J. Stat. Mech. P06002 (2004).


\bibitem{CalabreseCardyTimeDependentEnt}
P. Calabrese and J. Cardy, J. Stat. Mech. P04010 (2005).

\bibitem{Bravyi06}
S. Bravyi, M. B. Hastings and F. Verstraete, Phys. Rev. Lett. {\bf 97}, 050401 (2006).
\bibitem{Cramer08}
M. Cramer, C. M. Dawson, J. Eisert and T. J. Osborne
Phys. Rev. Lett. {\bf 100}, 030602 (2008).


\bibitem{EisertOsborne} J. Eisert and T. J. Osborne, Phys. Rev. Lett. {\bf 97}, 150404 (2006).

\bibitem{Bennett96}
C. H. Bennett, H. J. Bernstein, S. Popescu, B. Schumacher, Phys. Rev. A{\bf 53}, 2046 (1996).

\bibitem{Verstraete04}
F. Verstraete, D. Porras, and J. I. Cirac,
Phys. Rev. Lett. {\bf 93}, 227205 (2004).

\bibitem{Kitaev Preskill}
A. Kitaev and J. Preskill, Phys. Rev. Lett. {\bf 96}, 110404 (2006).
 
\bibitem{Levin Wen}
M. Levin and X. G. Wen, Phys. Rev. Lett. {\bf 96}, 110405 (2006).

\bibitem{Fradkin}
S. Dong, E. Fradkin, R. G. Leigh and S. Nowling, 
JHEP0805(2008)016; arXiv:0802.3231.

\bibitem{KlichLevitov08a} I. Klich, L. S. Levitov, ``Quantum Noise as an Entanglement Meter,'' Phys. Rev. Lett. (in print); arXiv:0804.1377

\bibitem{van Wees}
B. J. van Wees, H. van Houten, C. W. J. Beenakker, J. G. Williamson, L. P. Kouwenhoven, D. van der Marel, C. T. Foxon,
Phys. Rev. Lett. {\bf 60}, 848 (1988).

\bibitem{Lesovik01} 
G. B. Lesovik, T. Martin, G. Blatter, Eur. Phys. J. B {\bf 24}, 287 (2001).

\bibitem{Chtchelkatchev02}
N. M. Chtchelkatchev, G. Blatter, G. B. Lesovik, T. Martin, Phys. Rev. B {\bf 66}, 161320 (2002)

\bibitem{Beenakker03}  C. W. J. Beenakker, C. Emary, M. Kindermann, J. L. van Velsen,
Phys. Rev. Lett. {\bf 91}, 147901 (2003).

\bibitem{Samuelsson03}
P. Samuelsson, E. V. Sukhorukov, and M. Buttiker, Phys. Rev.
Lett. {\bf 91}, 157002 (2003).

\bibitem{Walborn}
S. P. Walborn {\it et al.},
Nature {\bf 440}, 1022 (2006).


\bibitem{LevitovLesovik}
L. S. Levitov and G. B. Lesovik, JETP Lett. {\bf 58}, 230 (1993).







\bibitem{Reulet03}
B. Reulet, J. Senzier, and D. E. Prober, Phys. Rev. Lett. {\bf 91}, 196601 (2003).

\bibitem{BomzeReznikov}
Yu. Bomze, G. Gershon, D. Shovkun, L. S. Levitov, M. Reznikov,
Phys. Rev. Lett. {\bf 95}, 176601 (2005).

\bibitem{Gershon07}
G. Gershon, Yu. Bomze, E. V. Sukhorukov, M. Reznikov,
Phys. Rev. Lett. {\bf 101}, 016803 (2008); arXiv:0710.1852

\bibitem{Fujisawa}
T. Fujisawa, T. Hayashi, R. Tomita and Y. Hirayama, 
Science {\bf 312}, 1634 (2006).

\bibitem{Gustavsson}
S. Gustavsson, 
R. Leturcq, B. Simovic, R. Schleser, T. Ihn, P. Studerus, K. Ensslin,
Phys. Rev. Lett. {\bf 96}, 076605 (2006).






\bibitem{MuzukantskiiAdamov}
B. A. Muzykantskii and Y. Adamov, Phys. Rev. B{\bf 68}, 155304 (2003).



\bibitem{Avron Graf} J. E. Avron, S. Bachmann, G. M. Graf and I. Klich, Comm. Math. Phys. {\bf 280}, 807 (2008).

\bibitem{KlichRefaelSilva}
I. Klich, G. Refael and A. Silva, Phys. Rev.  A{\bf 74}, 032306
(2006).



\bibitem{Abanov Ivanov}
A. G. Abanov and D. A. Ivanov, Phys. Rev. Lett. {\bf 100}, 086602 (2008).

\bibitem{Klich06} I. Klich, J. Phys. A {\bf 39}, L85 (2006).

\bibitem{Ivanov93} 
D. A. Ivanov and L. S. Levitov, Pis'ma Zh. Eksp. Teor. Fiz. {\bf 58}, 450 (1993)
[Engl. transl. JETP Lett. {\bf 58}, 461 (1993)]. 

\bibitem{Levitov96} 
L. S. Levitov, H.-W. Lee, and G. B. Lesovik, J. Math. Phys. (N.Y.) {\bf 37}, 4845 (1996).

\bibitem{Ivanov97} 
D. A. Ivanov, H. W. Lee, and L. S. Levitov, Phys. Rev. B {\bf 56}, 6839 (1997).

\bibitem{Vanevic} 
M. Vanevi\'c, Yu. V. Nazarov, and W. Belzig, Phys. Rev. Lett. {\bf 99}, 076601 (2007).

\bibitem{Sherkunov}
Y. B. Sherkunov, A. Pratap, B. Muzykantskii, and N. d'Ambrumenil, Phys. Rev. Lett. {\bf 100}, 196601 (2008)



\bibitem{Peschel}
I. Peschel, J. Phys. A:  Math. Gen. {\bf 36}, L205 (2003).

\bibitem{Beenakker_dS/dt}
C. W. J. Beenakker, Proc. Int. School Phys. E. Fermi, Vol. 162 (IOS Press, Amsterdam, 2006); cond-mat/0508488








\bibitem{LevitovLesovik04}
L. S. Levitov, G. B. Lesovik, 
arXiv:cond-mat/9401004 

\bibitem{KlichLevitov08b} I. Klich, L. S. Levitov, arXiv:0812.0006
\end{thebibliography}
\end{document}